\documentclass[runningheads]{llncs}
\usepackage{graphicx}
\usepackage[misc,geometry]{ifsym}
\usepackage{graphicx}
\usepackage{multirow}
\usepackage{algorithm}
\usepackage{algorithmic}
\usepackage{makecell}
\usepackage{tabularx}
\usepackage{multirow}
\usepackage{amsmath}
\usepackage{xcolor}
\usepackage{blkarray, bigstrut}
\usepackage{hyperref}
\hypersetup{colorlinks=true,citecolor=blue}
\usepackage{amsfonts}
\usepackage{siunitx}
\usepackage{dsfont}
\usepackage{lipsum}
\usepackage{adjustbox}
\usepackage{float}
\usepackage{booktabs}
\usepackage{cite}
\usepackage{bm}
\usepackage{wrapfig}
\usepackage{pifont}
\usepackage{relsize}
\usepackage{colortbl}
\definecolor{lightergray}{gray}{0.2}
\newcommand\eg {{\it e.g., }}

\newcommand\etc {{etc.}}

\newcommand\ie {{\it i.e., }}

\begin{document}
%

\title{
Mammo-CLIP: A Vision Language Foundation Model to Enhance Data Efficiency and Robustness in Mammography
}

\author{Shantanu Ghosh\inst{1}{$^{(\textrm{\Letter})}$}\and
Clare B. Poynton\inst{2}\and
Shyam Visweswaran\inst{3}\and
Kayhan Batmanghelich\inst{1}}
\authorrunning{S. Ghosh et al.}
%
\institute{
Department of Electrical and Computer Engineering, Boston University, Boston, MA, USA \\
\email{shawn24@bu.edu}\\
\and
Boston University Chobanian \& Avedisian School of Medicine, Boston, MA, USA \\
\and
Department of Biomedical Informatics, University of Pittsburgh, Pittsburgh, PA, USA 
}
\maketitle              
\begin{abstract}
The lack of large and diverse training data on Computer-Aided Diagnosis (CAD) in breast cancer detection has been one of the concerns that impedes the adoption of the system. 
Recently, pre-training with large-scale image text datasets via Vision-Language models (VLM) (\eg CLIP) partially addresses the issue of robustness and data efficiency in computer vision (CV). 
This paper proposes Mammo-CLIP, the first VLM pre-trained on a substantial amount of screening mammogram-report pairs, addressing the challenges of dataset diversity and size. Our experiments on two public datasets demonstrate strong performance in classifying and localizing various mammographic attributes crucial for breast cancer detection, showcasing data efficiency and robustness similar to CLIP in CV. We also propose Mammo-FActOR, a novel feature attribution method, to provide spatial interpretation of representation with sentence-level granularity within mammography reports. Code is available publicly: \url{https://github.com/batmanlab/Mammo-CLIP}.

\end{abstract}
\section{Introduction}
Breast cancer remains a leading cause of death among women, with a global death toll of 670K in 2022. Creating AI tools for cancer in mammograms has been a central focus for CAD~\cite{alberdi2005use, srivastava2013design, srivastava2014quantitative}. 
Yet, training robust CAD models requires a large and diverse dataset, which is expensive due to the high cost of collecting large-scale annotations.
VLMs (\eg CLIP~\cite{radford2021learning}) pre-trained on large image-text pairs partially improve robustness in CV. 
This paper introduces Mammo-CLIP, the first VLM trained with mammogram-report pairs.
Also, we propose Mammo-FActOR: \underline{\textbf{Mammo}}grams for \underline{\textbf{F}}eature \underline{\textbf{A}}ttribution \underline{\textbf{C}}onnec\underline{\textbf{T}}ing \underline{\textbf{O}}bservations and \underline{\textbf{R}}eports, 
novel feature attribution method, enhancing interpretability by finding textual attribute-aligned visual representations.

\noindent VLMs align image-text representations in joint embedding space, providing significant benefits. Pretraining on large datasets enhances generalizability~\cite{eslami2021does, liu2023clip} and data-efficiency~\cite{radford2021learning, li2022supervision} for downstream tasks. The language encoder reduces reliance on costly attribute annotations for developing interpretable models~\cite{yang2023language, oikarinen2023label}. In medical imaging, radiology reports are a source of weak labels, aiding disease localization~\cite{yu2022anatomy, muller2023anatomy}. Generalizability, interpretability, data efficiency, and weak supervision are crucial to improving diagnostic precision and insightful analysis in medical imaging.

\noindent Application of VLM in medical imaging has been mostly limited in chest X-ray (CXR) domain~\cite{zhang2022convirt, huang2021gloria, wu2023medklip, wang2022medclip, you2023cxr} due to availability of image-report dataset \eg MIMIC-CXR~\cite{johnson2019mimic}. VLMs in other domains are mostly created with images and captions from PubMed or public forums~\cite{kim2023fostering, lin2023pmc, huang2023visual}. 
Although fine-tuning the CLIP model on mammograms remains an option, fine-tinning reduces the image resolution to be consistent with the input dimension of CLIP. Reducing the resolution results in the loss of critical semantic visual cues essential for accurate interpretation and diagnosis. 
Also, identifying mammographic attributes requires the interpretation of high-resolution images, which CLIP does not offer.

\noindent\textbf{Contribution.} Our contributions are two fold: (1) We present Mammo-CLIP, the first VLM in mammography; (2) We introduce Mammo-FActOR, a novel feature attribution method, to align visual representations with attributes in mammography reports. Mammo-CLIP leverages a dataset of screening mammogram-report pairs. This dataset is collected by the University of Pittsburgh Medical Center (UPMC). Mammo-CLIP employs two key approaches for enhancing its learning capability:
1) a data augmentation strategy to expand the training data by incorporating external datasets containing images and mammographic attributes (\eg mass, calcification \etc) but lacking reports. We synthesize reports based on these attributes.
2) multi-view supervision (MVS)~\cite{li2022supervision}, to learn rich representations from limited data.
Our experiments on various breast cancer detection and localization tasks, conducted across two publicly available datasets, exhibit data efficiency and robustness similar to those observed in the CV community. Although Mammo-CLIP is trained on the screening dataset, where the cancer is not explicitly mentioned, fine-tuning its representation can bootstrap data-scare classification and abnormality localization experiments in completely out-of-training distributions.
The final contribution, Mammo-FActOR, successfully localizes the attributes without relying on the ground truth bounding boxes.

\label{sec:intro}

\section{Method}

\begin{figure*}[t]
\begin{center}
\includegraphics[width=\linewidth]{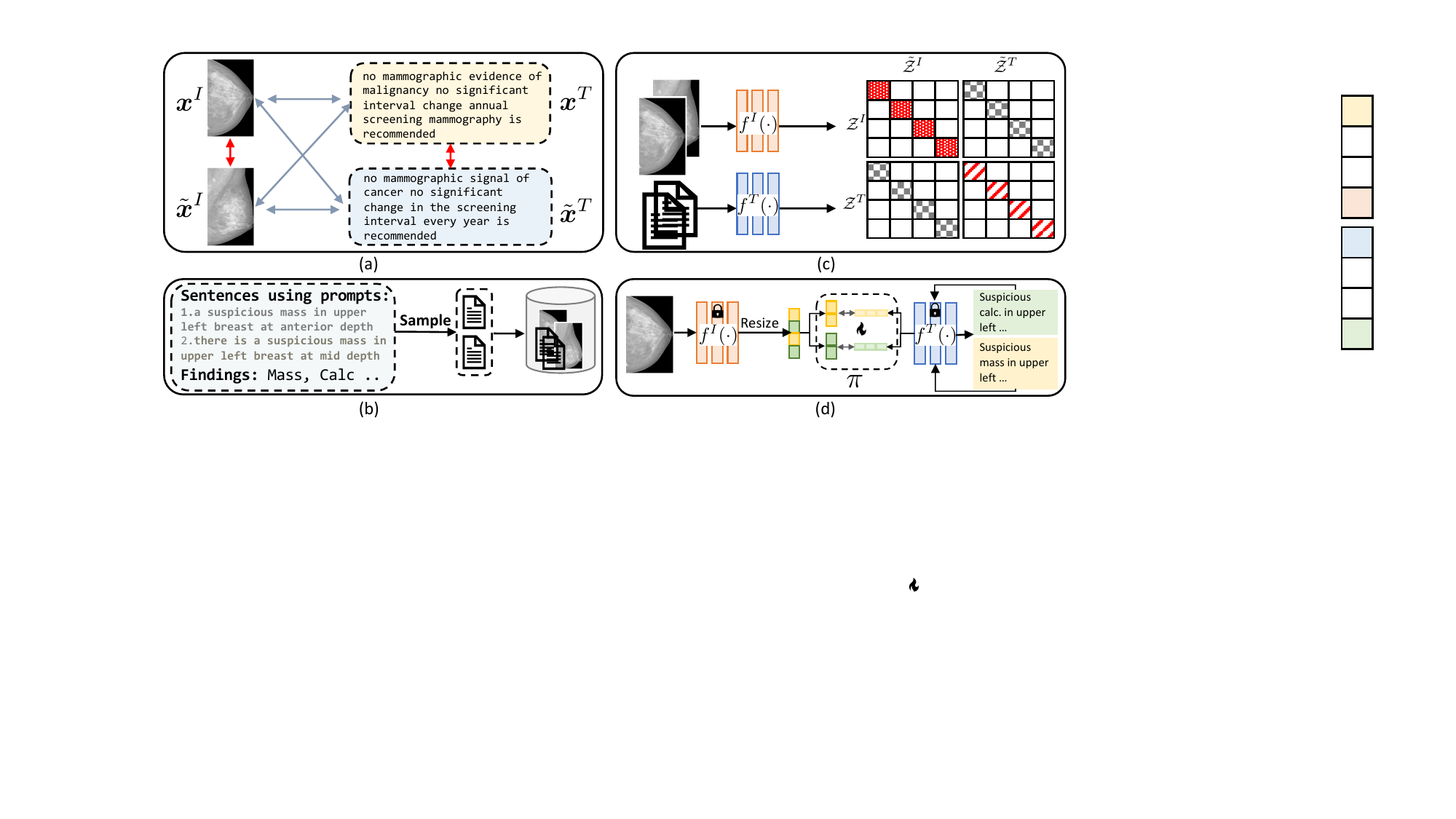}
\caption{Schematic view of our method. (a) 
Image-text augmentation for MVS.
(b) Dataset augmentation by synthesizing reports using image-attribute datasets. (c) Mammo-CLIP pretraining strategy. (d) Feature attribution using Mammo-FACtoR.}
\label{fig:schematic}
\end{center}
\end{figure*}


The goal is to align image ($\boldsymbol{x}^I$) and text ($\boldsymbol{x}^T$) representations via image ($f^I(\cdot)$) and text encoders ($f^T(\cdot)$). 
Our method consists of 3 components:
1) To enhance generalizability from limited data, Mammo-CLIP is based on Multi-View Supervision (MVS)~\cite{li2022supervision}. MVS leverages cross-modal self-supervision across original and augmented image-text pairs.
2) We use both instance and dataset augmentation strategies to bootstrap the generalizability further. 
3) Finally, we propose Mammo-FActOR, a novel interpretable module to map the visual representation to the textual attributes. 
We discuss the architectures of each component in detail in the experiment section.
Fig~\ref{fig:schematic} illustrates our approach.

\noindent \textbf{Notation.} Let $(\boldsymbol{x}_i^I, \boldsymbol{x}_i^T)$ and $(\Tilde{\boldsymbol{x}}_i^I, \Tilde{\boldsymbol{x}}_i^T)$ be the original and augmented image-text pair of patient $i$ respectively. 
Let $\mathcal{Z}^I = \{\boldsymbol{z^I}_i\}_{i=1}^B$ and $\mathcal{Z}^T = \{\boldsymbol{z^T}_i\}_{i=1}^B$ be the normalized representation of the image and text respectively of a batch of size $B$. Similarly, $\Tilde{\mathcal{Z}}^I$ and $\Tilde{\mathcal{Z}}^T$ denote the representation of the augmented image-text pairs.  We obtain the normalized representations by projecting the outputs from Mammo-CLIP's image and text encoders, $f^I(\cdot)$ and $f^T(\cdot)$, to the same dimensions, followed by $\ell_2$ normalization.

\subsection{Mammo-CLIP}
Given a batch of representation pairs $\mathcal{Z}, \Tilde{\mathcal{Z}}$, the following contrastive loss aligns the two representations
by pulling paired representations closer while pushing unpaired ones,
\begin{equation}
\label{equ: loss_INFO_NCE}
\mathcal{L}(\mathcal{Z}, \Tilde{\mathcal{Z}}) = -
     \sum_{\boldsymbol{z}\in \mathcal{Z}}\log \frac{\text{exp}(\langle\boldsymbol{z}, \Tilde{\boldsymbol{z}} \rangle/\tau)}{\sum_{\Tilde{\boldsymbol{z}} \in \Tilde{\mathcal{Z}}} \text{exp}(\langle\boldsymbol{z}, \Tilde{\boldsymbol{z}}\rangle/\tau)}
 ,
\end{equation}
\noindent where $\langle\cdot,\cdot\rangle$ and $\tau$ are the similarity function and a learnable temperature to scale the logits, respectively. Mammo-CLIP utilizes self-supervision across every pair of representations using the following MVS loss,



\begin{equation}
\label{equ:loss_INFO_NCE}
\mathcal{L} =
     \sum_{\mathcal{Z}, \tilde{\mathcal{Z}}\in \{\mathcal{Z}^I, \tilde{\mathcal{Z}}^I, \mathcal{Z}^T, \tilde{\mathcal{Z}}^T\}, \mathcal{Z} \neq \tilde{\mathcal{Z}}}
     \mathcal{L}(\mathcal{Z}, \tilde{\mathcal{Z}}).
\end{equation}
In our experiments, we down-weight $\mathcal{L}(\mathcal{Z}^T, \Tilde{\mathcal{Z}}^T)$, the loss for the original and augmented text representation pair by half.



\subsection{Instance and Dataset Augmentation}
\label{sec:dataset_aug}
\noindent\textbf{Instance augmentation.}
For $i^{th}$ patient, $\boldsymbol{x}^I_i$ and $\tilde{\boldsymbol{x}}^I_i$ denote the CC and MLO views respectively. We also use affine and elastic transformations to extend the augmented sets.
Likewise, $\boldsymbol{x}^T_i$ and $\tilde{\boldsymbol{x}}^T_i$ denote the impression and findings sections of the mammography reports. 
We followed the technique proposed in~\cite{you2023cxr} that uses translation between two languages to generate augmented text.

\noindent\textbf{Dataset augmentation.} 
In addition to the dataset containing paired mammograms and reports, we leverage a separate dataset containing mammograms and attributes (\eg \textit{mass, calcification}) to mitigate the challenge of the limited dataset further. To synthesize reports from the attribute, we ask a board-certified radiologist to construct a set of prompts using the values of the attributes (\textit{positive}, \textit{negative} \etc), subtypes of the attribute (\eg \textit{suspicious}, \textit{obscured} \etc), laterality of the breast (\ie \textit{right} and \textit{left}), depth of the breast (\ie \textit{anterior}, \textit{mid}, \textit{posterior}) and position of the breast (\eg \textit{upper}, \textit{lower} \etc). We synthesize the report by substituting the attribute values for each patient to a randomly selected prompt (see Fig. 1 in the Appendix). 


\begin{table*}[t]
\caption{Classification performance on VinDr dataset. 
We report Area Under the Curve (AUC) AUC for binary classification tasks to classify calcification and mass. We report accuracy for multi-class classification tasks to predict density.
}
\begin{adjustbox}{max width=0.95\textwidth, center}
\begin{tabular}{lccccccc}
\toprule
\multicolumn{1}{c}{} &
  \multicolumn{1}{c}{} &
  \multicolumn{1}{c}{} &
  \multicolumn{1}{c}{\textbf{Zero-shot (ZS)}} & 
  \multicolumn{3}{c}{\textbf{Linear Probe (LP)}} &
  \multicolumn{1}{c}{\textbf{Finetune (FT)}} \\ 
\cmidrule(lr){4-4}
\cmidrule(lr){5-7}
\cmidrule(lr){8-8}
\multicolumn{1}{l}{\multirow{-2}{*}{\textbf{Label}}} &
  \multicolumn{1}{c}{\multirow{-2}{*}{\textbf{Model}}} &
  \multicolumn{1}{c}{\multirow{-2}{*}{\textbf{Pre-training Data}}} &
  \multicolumn{1}{c}{\scriptsize 100\%} & 
  \multicolumn{1}{c}{\scriptsize 10\%} &
  \multicolumn{1}{c}{\scriptsize 50\%} &
  \multicolumn{1}{c}{\scriptsize 100\%} &
  \multicolumn{1}{c}{\scriptsize 100\%} \\ 
\midrule
 &
  CLIP w/ RN-50 &
  UPMC &
  $0.37$ &
  $0.42$ &
  $0.51$ &
  $0.55$ &
  $0.68$ \\
 &
  CLIP w/ EN-B5 &
  UPMC &
  $0.53$ &
  $0.53$ &
  $0.60$ &
  $0.65$ &
  $0.90$ \\
&
  Mammo-CLIP w/ EN-B2  &
  UPMC &
  $\boldsymbol{0.68}$ &
  $0.90$ &
  $0.92$ &
  $0.92$ &
  $\bm{0.98}$ \\
&
  Mammo-CLIP w/ EN-B5 &
  UPMC &
  $0.61$ &
  \underline{$0.91$} &
  \underline{$0.93$} &
  \underline{$0.94$} &
  $\bm{0.98}$ \\
\multirow{-5}{*}{Calcification} &
  \cellcolor{lightgray}Mammo-CLIP w/ EN-B5 &
  \cellcolor{lightgray}UPMC, VinDr &
  \cellcolor{lightgray}$0.62$ &
  \cellcolor{lightgray}$\bm{0.92}$ &
  \cellcolor{lightgray}$\bm{0.94}$ &
  \cellcolor{lightgray}$\bm{0.96}$ &
  \cellcolor{lightgray}$\bm{0.98}$ \\
\midrule
&
  CLIP w/ RN-50 &
  UPMC &
  $0.20$ &
  $0.34$ &
  $0.42$ &
  $0.49$ &
  $0.57$ \\
 &
  CLIP w/ EN-B5 &
  UPMC &
  $0.32$ &
  $0.52$ &
  $0.57$ &
  $0.59$ &
  $0.78$ \\
&
  Mammo-CLIP w/ EN-B2 &
  UPMC &
  \underline{$0.58$} &
  $0.69$ &
  $0.72$ &
  $0.73$ &
  \underline{$0.85$} \\
&
  Mammo-CLIP w/ EN-B5 &
  UPMC &
  $0.48$ &
  \underline{$0.73$} &
  \underline{$0.78$} &
  \underline{$0.79$} &
  \underline{$0.85$} \\
\multirow{-6}{*}{Mass} &
  \cellcolor{lightgray}Mammo-CLIP w/ EN-B5 &
  \cellcolor{lightgray}UPMC, VinDr &
  \cellcolor{lightgray}$\bm{0.76}$ &
  \cellcolor{lightgray}$\bm{0.80}$ &
  \cellcolor{lightgray}$\bm{0.84}$ &
  \cellcolor{lightgray}$\bm{0.86}$ &
  \cellcolor{lightgray}$\bm{0.88}$ \\
\midrule
&
  CLIP w/ RN-50 &
  UPMC &
  $0.04$ &
  $0.55$ &
  $0.61$ &
  $0.69$ &
  $0.72$ \\
 &
  CLIP w/ EN-B5 &
  UPMC &
  $0.10$ &
  $0.76$ &
  $0.76$ &
  $0.78$ &
  $0.83$ \\
&
  Mammo-CLIP w/ EN-B2 &
  UPMC &
  $0.13$ &
  \underline{$0.80$} &
  $0.82$ &
  $0.84$ &
  \underline{$0.85$} \\
&
  Mammo-CLIP w/ EN-B5 &
  UPMC &
  $0.15$ &
  $\bm{0.83}$ &
  \underline{$0.84$} &
  \underline{$0.85$} &
  $\bm{0.88}$ \\
\multirow{-6}{*}{Density} &
  \cellcolor{lightgray}Mammo-CLIP w/ EN-B5 &
  \cellcolor{lightgray}UPMC, VinDr &
  \cellcolor{lightgray}$0.15$ &
  \cellcolor{lightgray}$\bm{0.83}$ &
  \cellcolor{lightgray}$\bm{0.86}$ &
  \cellcolor{lightgray}$\bm{0.86}$ &
  \cellcolor{lightgray}$\bm{0.88}$ \\
\bottomrule
\end{tabular}
\end{adjustbox}
\label{table:classification_vindr_cls}
\end{table*}

\subsection{Mammo-FActOR}
Attributes mentioned in the radiology report can be viewed as weak labels. The Mammon-CLIP learns to align image and text representation at the global level. To obtain a spatial interpretation of the learned representation at the granularity of the attribute, we develop Mammo-FActOR. The Mammo-FActOR learns to align sentences in the radiology reports containing an attribute to the channels of the frozen image encoder, $f^I(\cdot)$. Let's assume we have $K$ different attributes and  $f^I(\mathbf{x}_i^{I}) \in \mathbb{R}^{C \times H \times W}$ is the output of the frozen image encoder and $\mathbf{t}_i^k \in \mathbb{R}^d$ is the representation of the sentence containing the $k$'th attribute in the report of the patient $i$. We use $\mathcal{X}_{-}^k$ to denote all mammography images not containing the $k$'s attribute. For attribute $k$, the Mammo-FActOR learns a mapping $\pi^k(\mathbf{t}_i^k, \mathbf{x}_i^{I} ) = \left( \text{MLP}( f^I(\mathbf{x}_i^{I}); \theta_k ) \mathbf{t}_i^k \right) \in \mathbb{R}^C$  representing the similarity of the channels with $\mathbf{t}_i^k$. The $\text{MLP}(\cdot; \theta_k)$ is a Multi-layer Perception parametrized by $\theta_k$. To learn the parameters of the MLP, we minimize the contrastive loss between images with and without attributes:
\begin{equation}
\mathcal{L}^{\text{fac}} = - \sum_{i,c,k} {\log \frac{ \exp\left( \left[ \pi^k(\mathbf{t}_i^k, \mathbf{x}_i^{I} ) \right]_c /\tau\right)}{  \exp\left( \left[ \pi^k(\mathbf{t}_i^k, \mathbf{x}_i^{I} ) \right]_c /\tau \right) + \sum_{x \in \mathcal{X}_{-}^k} { {  \exp\left( \left[ \pi^k(\mathbf{t}_i^k, \mathbf{x} ) \right]_c /\tau\right) } } }   },
\end{equation}
where $\left[\cdot\right]_c$ denotes $c$'th element of the input vector argument and $\tau$ is the temperature.
After training, we weigh the representations as, $\pi^k(\cdot)f^I(\cdot) \in \mathbb{R}^{H\times W}$ to construct the textual aligned heatmap for attribute $k$.
\label{sec:methods}

\section{Experiments}
We aim to answer the following research questions:
\textbf{RQ1.} Does Mammo-CLIP enhance zero-shot (ZS) capabilities and labeling efficiency in classification tasks? 
\textbf{RQ2.} Does Mammo-CLIP learn robust representations?
\textbf{RQ3.} Does Mammo-CLIP boost the performance of label-efficient localization?
\textbf{RQ4.} Does Mammo-FActOR enhance interpretability through attribute-aligned representation?

\begin{table*}[t]
\caption{Classification performance on RSNA dataset to classify malignancy.}
\begin{adjustbox}{max width=0.95\textwidth, center}
\begin{tabular}{ccccccc}
\toprule
\multicolumn{1}{c}{} &
  \multicolumn{1}{c}{} &
  \multicolumn{1}{c}{\textbf{Zero-shot (ZS)}} & 
  \multicolumn{3}{c}{\textbf{Finetune (FT)}} &
  \multicolumn{1}{c}{\textbf{Linear Probe (LP)}} \\ 
\cmidrule(lr){3-3}
\cmidrule(lr){4-6}
\cmidrule(lr){7-7}
  \multicolumn{1}{c}{\multirow{-2}{*}{\textbf{Model}}} &
  \multicolumn{1}{c}{\multirow{-2}{*}{\textbf{Pre-training Data}}} &
  \multicolumn{1}{c}{\scriptsize 100\%} & 
  \multicolumn{1}{c}{\scriptsize 10\%} &
  \multicolumn{1}{c}{\scriptsize 50\%} &
  \multicolumn{1}{c}{\scriptsize 100\%} &
  \multicolumn{1}{c}{\scriptsize 100\%} \\ 
\midrule
CLIP w/ RN-50 &
  UPMC &
  $0.28$ &
  $0.62$ &
  $0.67$ &
  $0.72$ &
  $0.58$ \\
CLIP w/ EN-B5 &
  UPMC &
  $0.42$ &
  $0.80$ &
  $0.84$ &
  $0.86$ &
  $0.70$ \\
Mammo-CLIP w/ EN-B2 &
  UPMC &
  $\underline{0.60}$ &
  $\underline{0.82}$ &
  $0.86$ &
  $0.89$ &
  $0.72$ \\
Mammo-CLIP w/ EN-B5 &
  UPMC &
  $\bm{0.62}$ &
  $\bm{0.85}$ &
  $\underline{0.89}$ &
  $\underline{0.90}$ &
  $\underline{0.75}$ \\
\cellcolor{lightgray}Mammo-CLIP w/ EN-B5 &
  \cellcolor{lightgray}UPMC, VinDr &
  \cellcolor{lightgray}\underline{$0.60$} &
  \cellcolor{lightgray}$\bm{0.85}$ &
  \cellcolor{lightgray}$\bm{0.90}$ &
  \cellcolor{lightgray}$\bm{0.91}$ &
  \cellcolor{lightgray}$\bm{0.79}$ \\
\bottomrule
\end{tabular}
\end{adjustbox}
\label{table:classification_rsna}
\end{table*}

\subsection{Datasets}
\textbf{UPMC.} 
The UPMC dataset includes 13,829 patient-report pairs, resulting in 25,355 screening mammograms (BI-RADS 0-2) with patients having at least one CC or MLO view. We use 0.8/0.2 for the train/val split.
\noindent\textbf{VinDr.}
The VinDr-Mammo~\cite{nguyen2023vindr} is a publicly available dataset comprising 5,000 exams (20,000 images) from Vietnam, each with four views, breast level BI-RADS assessment category (1-5), breast density category (A-D), annotating and location of mammographic attributes (\ie mass, calcifications, \etc). We use the train-test split in~\cite{nguyen2023vindr}.
\noindent\textbf{RSNA.}
The RSNA-Mammo\footnote{\url{https://www.kaggle.com/competitions/rsna-breast-cancer-detection}} is a publicly available dataset having 11913 patients with 486 cancer cases. We use train/val/test split as 0.7/0.2/0.1.

\subsection{Experimental details}
\noindent \textbf{Pre-processing the images.} We use a rule-based approach to find the breast's ROI for all the datasets. We set values less than 40 to 0 and eliminate the consistently identical rows and columns, as they denote the background. It results in images with an average aspect ratio of 1:1.6 to 2, finally resized to $1520\times912$.

\noindent\textbf{Pre-training.} We develop 2 variant of Mammo-CLIP models: one pre-trained with the UPMC dataset only and the other using both the UPMC and VinDr datasets. Mammo-CLIP utilizes BioClinicalBERT~\cite{alsentzer2019publicly} as the text encoder and EfficientNet(EN)-B2 and B5~\cite{tan2019efficientnet} with ImageNet pre-trained weights as the image encoder. Image augmentations consist of 1) affine transformations with rotations up to $20^\circ$, a minimum translation of 0.1\%, scaling factors [0.8, 1.2], and shearing by $20^\circ$; and 2) elastic transformations with $(\alpha=10$, $\sigma=5)$. Text augmentations include sentence swaps and back-translation from Italian to English\footnote{\url{https://huggingface.co/Helsinki-NLP}}. We train Mammo-CLIP for 10 epochs
in a mixed-precision manner and optimize
by AdamW~\cite{loshchilov2017decoupled} with an initial learning rate 5e-5 and a weight decay 1e-4. We
use a cosine-annealing learning-rate scheduler~\cite{loshchilov2016sgdr} with a warm-up for 1 epoch.
\noindent\textbf{Baseline.} Using UPMC dataset and CLIP objective~\cite{radford2021learning}, we construct two baselines: 1) an image encoder w/ ResNet (RN)-50 initialized with CLIP weights and fine-tuned with $224\times224$ images, 2) EfficientNet (EN)-B5 fine-tuned using the same pre-processed images as Mammo-CLIP. Both the baselines are pre-trained using the UPMC dataset, as CLIP only uses an image-text dataset, not an image-attribute dataset.
\textbf{Mammo-FActOR.} Our Mammo-Factor is the same light-weight neural network~\cite{varma2023villa} with linear layer$\rightarrow$ReLU$\rightarrow$linear layer. We train it with a learning rate 0.0001 and temperature ($\tau$) of 0.007 for 20 epochs.
\begin{table*}[t]
\caption{Localization performance (mAP) on VinDr dataset.}
\begin{adjustbox}{max width=0.95\textwidth, center}
\begin{tabular}{lcccccc}
\toprule
\multicolumn{1}{c}{} &
  \multicolumn{1}{c}{} &
  \multicolumn{1}{c}{} &
  \multicolumn{3}{c}{\textbf{Finetune}} &
  \multicolumn{1}{c}{\textbf{Freeze Encoder}} \\ 
\multicolumn{1}{l}{\multirow{-2}{*}{\textbf{Label}}} &
  \multicolumn{1}{c}{\multirow{-2}{*}{\textbf{Model}}} &
  \multicolumn{1}{c}{\multirow{-2}{*}{\textbf{Pre-training Data}}} &
  \multicolumn{1}{c}{\scriptsize 10\%} &
  \multicolumn{1}{c}{\scriptsize 50\%} &
  \multicolumn{1}{c}{\scriptsize 100\%} &
  \multicolumn{1}{c}{\scriptsize 100\%} \\ 
\midrule
 &
  CLIP w/ RN-50 &
  UPMC &
  $0.03$ &
  $0.11$ &
  $0.17$ &
  $0.08$ \\
 &
  CLIP w/ EN-B5 &
  UPMC &
  $0.09$ &
  $0.23$ &
  $0.29$ &
  $0.10$ \\
&
  Mammo-CLIP w/ EN-B2  &
  UPMC &
  $0.08$ &
  $0.20$ &
  \underline{$0.32$} &
  $\bm{0.17}$ \\
&
  Mammo-CLIP w/ EN-B5 &
  UPMC &
  $\bm{0.12}$ &
  $\bm{0.28}$ &
  \bm{$0.35$} &
  $\bm{0.17}$ \\
\multirow{-5}{*}{Calcification} &
  \cellcolor{lightgray}Mammo-CLIP w/ EN-B5 &
  \cellcolor{lightgray}UPMC, VinDr &
  \cellcolor{lightgray}\underline{$0.10$} &
  \cellcolor{lightgray}\underline{$0.25$} &
  \cellcolor{lightgray}\bm{$0.35$} &
  \cellcolor{lightgray}$\bm{0.17}$ \\
\midrule
&
  CLIP w/ RN-50 &
  UPMC &
  $0.23$ &
  $0.29$ &
  $0.36$ &
  $0.15$ \\
 &
  CLIP w/ EN-B5 &
  UPMC &
  $0.34$ &
  $0.45$ &
  $0.49$ &
  $0.28$ \\
&
  Mammo-CLIP w/ EN-B2  &
  UPMC &
  $0.38$ &
  $0.50$ &
  \underline{$0.55$} &
  \underline{$0.37$} \\
&
  Mammo-CLIP w/ EN-B5 &
  UPMC &
  \underline{$0.41$} &
  \underline{$0.52$} &
  $\bm{0.58}$ &
  $0.35$ \\
\multirow{-5}{*}{Mass} &
  \cellcolor{lightgray}Mammo-CLIP w/ EN-B5 &
  \cellcolor{lightgray}UPMC, VinDr &
  \cellcolor{lightgray}$\bm{0.43}$ &
  \cellcolor{lightgray}$\bm{0.54}$ &
  \cellcolor{lightgray}$\bm{0.58}$ &
  \cellcolor{lightgray}$\bm{0.39}$ \\
\bottomrule
\end{tabular}
\end{adjustbox}
\label{table:classification_vindr_obj}
\end{table*}

\begin{figure*}[h]
\begin{center}
\includegraphics[width=\linewidth]{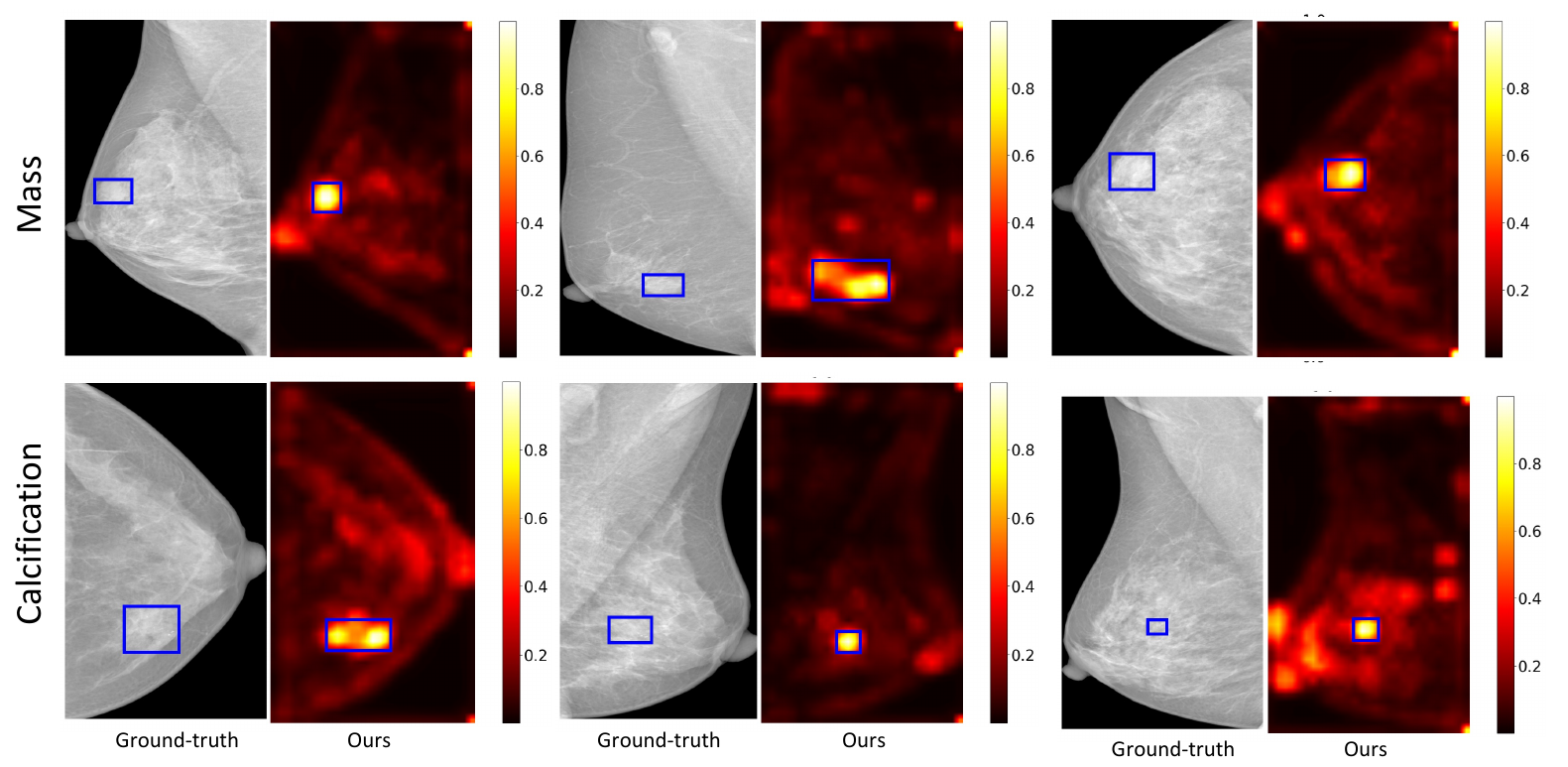}
\caption{Mammo-FACtoR localizes mass and calcification w/o the ground-truth bboxes from the VinDr dataset. For each pair, the left image denotes the original image with the ground-truth bbox, while the right one is the bbox predicted by Mammo-Factor.}
\label{fig:clip_facto}
\end{center}
\end{figure*}

\noindent \textbf{Downstream evaluation.} Evaluation of the downstream tasks utilizes both the Mammo-CLIP's image representations to 1) classify mass, calcification, and density in the VinDr dataset and cancer in the RSNA dataset; 2) localize mass and calcification in the VinDr dataset on a held-out test set. We evaluate the classification tasks in 3 settings: zero-shot (ZS), linear probe (LP), and fine-tuning (FT). In ZS and LP, we freeze both the encoders of Mammo-CLIP. For FT, we jointly fine-tune the image encoder of Mammo-CLIP and the linear classifier. For zero-shot (ZS) evaluation, we use prompts as \{No $\langle E \rangle$, $\langle E \rangle$\}, where $E$ denotes mass, calcification, and cancer. Refer to Tab. 1 in the Appendix for prompts for density. LP on VinDr utilizes varying amounts of the training data (10\%, 50\%, or 100\%). For FT on VinDr, we use 100\% data. As the screening mammogram reports do not mention malignancy explicitly, data-efficient cancer classification for the RSNA dataset involves fine-tuning with varying proportions of the training data (10\%, 50\%, or 100\%). We also report LP for the RSNA dataset using the entire training set. 
The localization task includes fine-tuning the image encoder of Mammo-CLIP using RetinaNet~\cite{lin2017focal} ($\alpha=0.6$, $\gamma=2.0$) model and 10\%, 50\%, and 100\% of training data from VinDR. We also report the localization performance of RetinaNet by freezing Mammo-CLIP's image encoder with 100\% of training data. Both setups in localization use only VinDr's suspicious findings. 
\textbf{Metric.} We report AUC scores for binary classification tasks to predict mass, calcification, and cancer and accuracy for multi-class classification tasks to predict density. For localization, we report mAP with IoU=0.5 and detector confidence thresholded to 0.05~\cite{lin2017focal} unless specified. 


\label{sec:experiment}

\section{Results}
\noindent \textbf{Classification (RQ1 \& RQ2).}
\label{classification}
\noindent Tab.~\ref{table:classification_vindr_cls} and~\ref{table:classification_rsna} illustrate the classification results for VinDr and RSNA, respectively. The Mammo-CLIP w/ EN-B5, pre-trained on the UPMC and VinDr datasets, outperforms all the other models in most of the evaluation settings. For VinDr, Mammo-CLIP's LP efficacy with 10\% of data surpasses the FT performance of the baselines with 100\% data. Also, Mammo-CLIP excels in cancer classification on the RSNA dataset, showing its robustness, even when cancer is not explicitly mentioned in the reports.

\begin{table*}[t]
\caption{Weakly supervised localization on VinDr using Mammo-FActOR. 
}
\centering
\begin{adjustbox}{max width=0.75\textwidth, center}
\begin{tabular}{ccccc}
\toprule
\multicolumn{1}{c}{\multirow{2}{*}{\textbf{Model}}} &
  \multicolumn{2}{c}{\textbf{Mass}} &
  \multicolumn{2}{c}{\textbf{Calcification}} \\ 
  \cmidrule(lr){2-3} 
  \cmidrule(lr){4-5}
\multicolumn{1}{c}{} & IoU@0.25 & IoU@0.50 & IoU@0.25 & IoU@0.50 \\ 
\cmidrule(lr){1-1}
\cmidrule(lr){2-3}
\cmidrule(lr){4-5}

  Mammo-CLIP w/ EN-B2 &
  $0.38$ &
  $0.22$ &
  $0.2$ &
  $0.13$ \\

  Mammo-CLIP w/ EN-B5 &
  \bm{$0.45$} &
  $0.37$ &
  \bm{$0.25$} &
  $0.17$ \\
  
  \bottomrule
\end{tabular}
\end{adjustbox}
\label{table:clip_factor}
\end{table*}

\noindent CLIP w/ RN-50 backbone fails to capture the visual cues from low-resolution images. CLIP w/ EN-B5 addresses this by employing high-resolution images but struggles to learn rich representation primarily due to its optimization strategy focusing solely on the contrastive loss of the original image-text pairs. High-resolution images and MVS optimization strategy allow Mammo-CLIP w/ EN backbones to extract richer representations from limited data than baselines, improving the classification performance. 
Recall that the UPMC dataset consists of mammograms from a different population than the VinDr dataset. The image encoder from Mammo-CLIP pre-trained with the UPMC dataset shows excellent performance on both LP and FT settings on the VinDR dataset, showcasing the robustness of the learned representations. ZS results on some tasks (especially density) exhibit the need for more data. 

\noindent \textbf{Localization (RQ2 \& RQ3).}
Tab.~\ref{table:classification_vindr_obj} shows localization results in a supervised setting, utilizing the bounding boxes of VinDr. Mammo-CLIP w/EN-B5 pre-trained with the UPMC and VinDr dataset outperforms other models in most settings. Refer to the \underline{Classification} subsection (described above) for insights into such improvement. Also, the utilization of VinDr during the pre-training improves the performance of both downstream tasks.

\noindent \textbf{Evaluating Mammo-FActOR (RQ4).} We evaluate the Mammo-FActOR using a weakly supervised localization task to localize the mass and calcification of the VinDr dataset without using the ground truth bounding boxes (bboxes). We construct heatmaps using the Mammo-FActOR for both attributes.
Following~\cite{yu2022anatomy}, we extract isolated regions from the heatmap where pixel values are greater than the 95\% quantile of the
heatmap’s pixel value distribution and generate bboxes.
We estimate mAP by comparing the IoU between generated and ground truth bounding boxes. A generated box is true positive when $\text{IoU} > T(\text{IoU})$, where $T(\cdot)$ is a threshold. Tab.~\ref{table:clip_factor} demonstrates the localization results. Mammo-CLIP w/ EN-B5 for $T(IoU) = 0.25$ outperforms the other variants for both the attributes. 
Fig.~\ref{fig:clip_facto} shows the qualitative results. Mammo-FActOR aligns image representations with attributes-specific textual descriptions from reports. Thus, it successfully localizes the attributes without relying on ground-truth bounding boxes during finetuning. Also, Fig. 2 in the Appendix shows the top 3 feature maps by Mammo-FActOR for detecting these attributes.

\label{sec:results}

\section{Conclusion}
This paper proposes a Mammo-CLIP and Mammo-FActOR to learn a cross-modal embedding, improving generalizability, data efficiency, and interpretability. 
Future directions include further leveraging vision transformers and cross-attention mechanisms to enhance Mammo-CLIP and Mammo-FActOR.
\label{sec:conclusion}

\bibliographystyle{splncs04}
\bibliography{main}

\section*{Supplementary materials}
\setcounter{figure}{0}
\setcounter{table}{0}
\subsection{Zero-shot prompts for density}
\begin{table}[H]
\caption{Zero-shot prompts for classifying density}
\label{tab:g_config_mimic_cxr}
\begin{center}
\begin{tabular}{c }
\toprule  
    \textbf{Prompts} \\
\midrule 
       1.the breasts being almost entirely fatty,\\ 2. scattered areas of fibroglandular density,\\ 3. the breast tissue is heterogeneously dense, \\ 4. the breasts are extremely dense \\
\bottomrule
\end{tabular}
\end{center}
\end{table}

\subsection{Prompts to synthesize report-like sentences from attributes in image-mammographic attribute dataset}
\begin{figure*}[h]
\begin{center}
\centerline{\includegraphics[height=7.5cm]{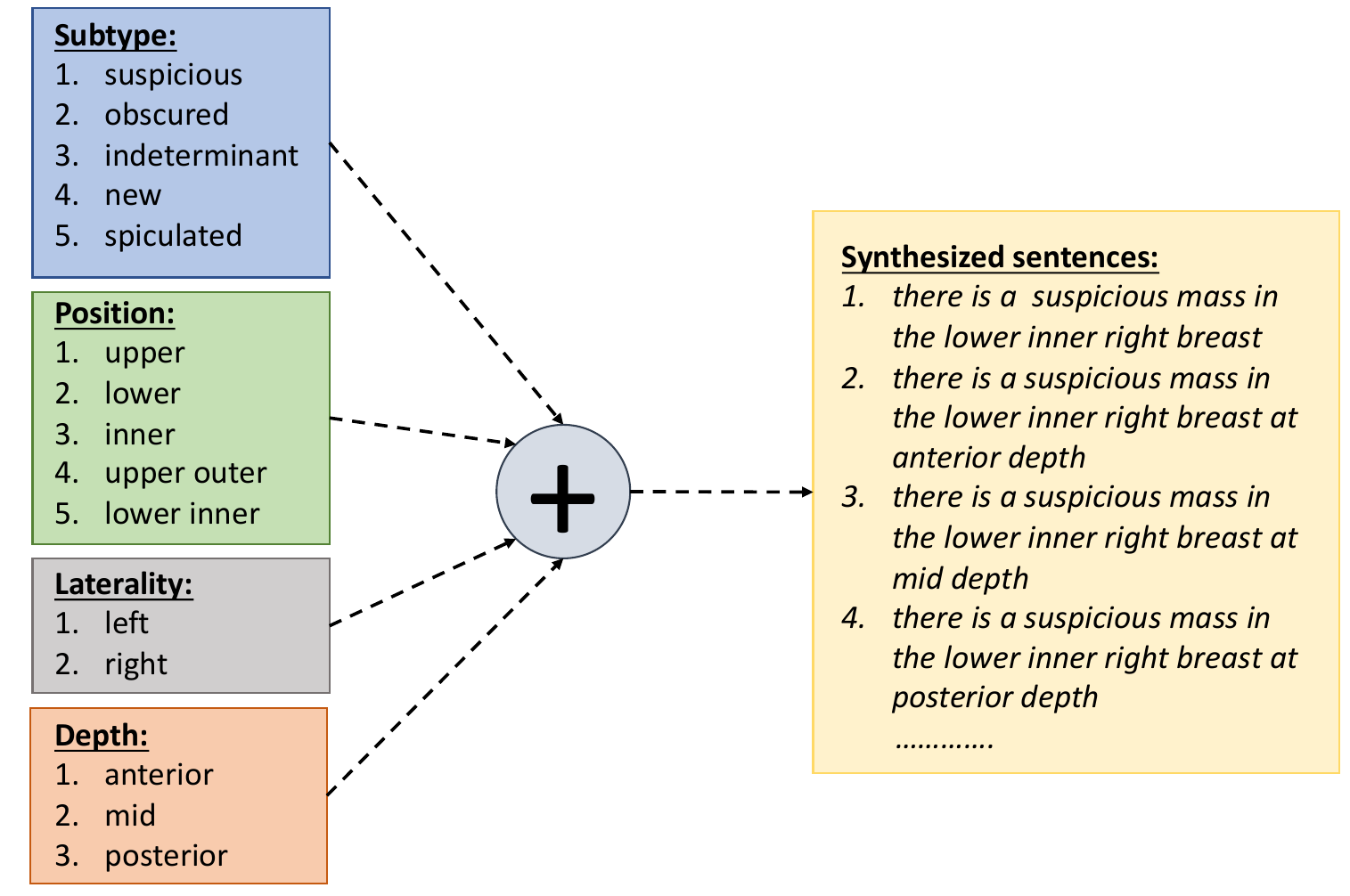}}
\caption{Example of report-like sentence generation for the attribute \textit{mass} labeled positively in the VinDr dataset using the subtypes of mass and position, laterality, and depth of the breast. We include all such prompts in our codebase in details.}
\label{fig:expert_performance_cv_vit}
\end{center}
\end{figure*}

\subsection{Visualization of activation maps identified by Mammo-FActOR}
\begin{figure*}[h]
\begin{center}
\centerline{\includegraphics[height=10cm]{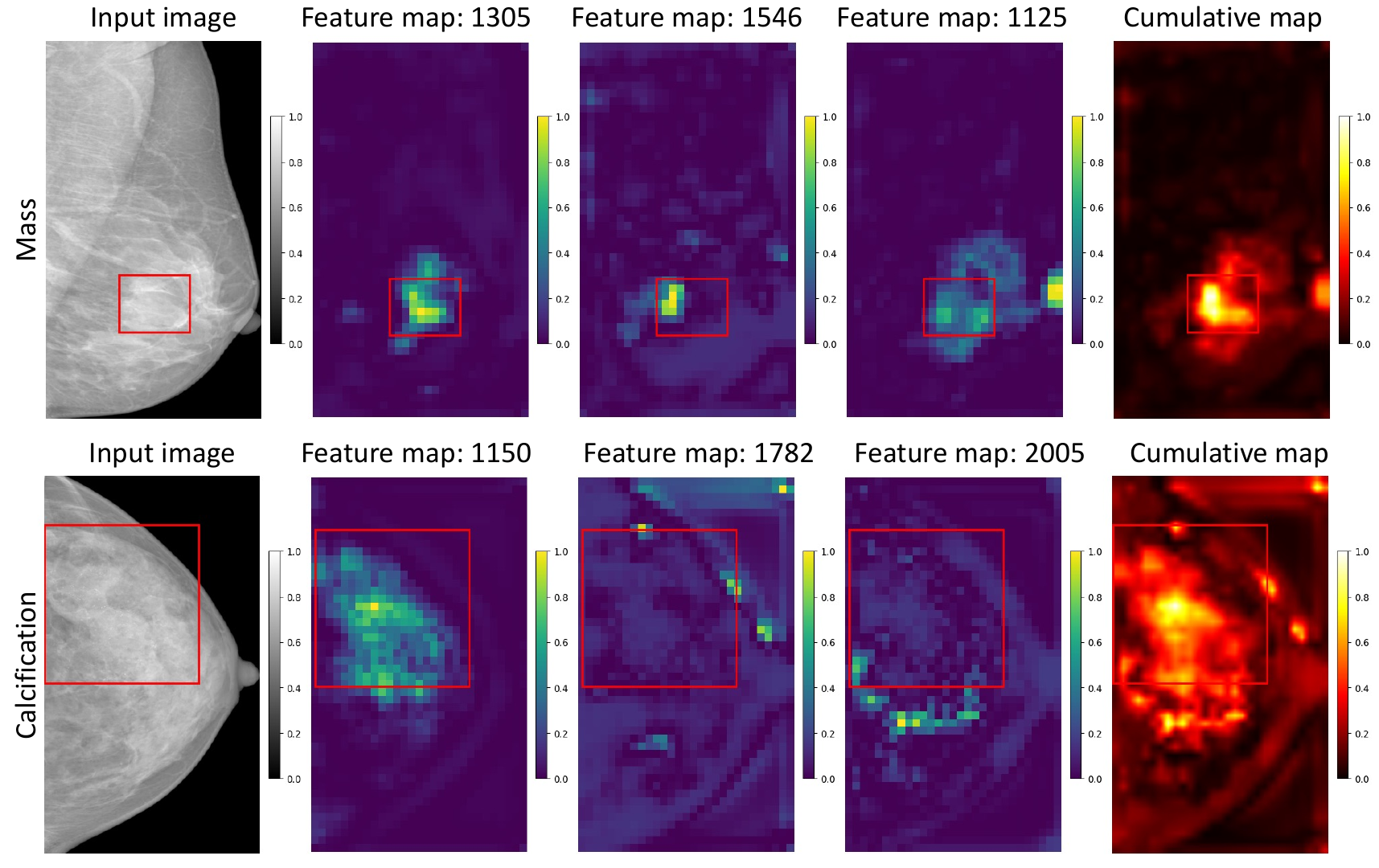}}
\caption{Visualization of activation maps identified by Mammo-FActOR to localize mass and calcification attributes via report. For each row, the left image is the original image. The middle three denote top-3 (left to right) activation maps with the respective unit number. The right one indicates the summation of the top-3 feature maps. We show the ground-truth bounding boxes with the red rectangle. Notably, the optimal feature map identified by Mammo-FActOR (Feature map unit 1305 for mass and 1150 for calcification) successfully localizes the attributes.}
\label{fig:expert_performance_cv_vit}
\end{center}
\end{figure*}

\label{sec:supplemental}
\end{document}